\documentclass[10pt]{article}

\usepackage{graphicx}
\usepackage{subcaption}

\usepackage[a4paper, left=35mm,right=35mm,top=34mm,bottom=34mm]{geometry}
\usepackage[utf8]{inputenc}
\usepackage[T1]{fontenc}
\usepackage[english]{babel}

\usepackage{enumerate}
\usepackage{graphicx}
\usepackage{hyperref}
\hypersetup{
    colorlinks=true,
    linkcolor=blue,
    filecolor=magenta,      
    urlcolor=cyan,
}

\usepackage{mathtools,amsthm,amssymb,amsfonts}
\usepackage{algorithm}
\usepackage{algpseudocode}
\makeatletter
\def\thm@space@setup{
  \thm@preskip=10pt \thm@postskip=10pt
}
\makeatother
\theoremstyle{plain}

\theoremstyle{plain}

\theoremstyle{definition}

\theoremstyle{definition}

\theoremstyle{remark}

\theoremstyle{remark}

\usepackage{caption} 
\usepackage{booktabs}
\usepackage{listings}
\usepackage{color}
\definecolor{dkgreen}{rgb}{0,0.6,0}
\definecolor{gray}{rgb}{0.5,0.5,0.5}
\definecolor{mauve}{rgb}{0.58,0,0.82}
\usepackage{enumitem}

\newcommand{\email}[1]{\protect\href{mailto:#1}{#1}}

\usepackage{xcolor}[2.11]
\colorlet{inlinkcolor}{green!50!black}
\colorlet{exlinkcolor}{red!50!black}
\if@hidelinks
\hypersetup{ hidelinks = true }
\else
\hypersetup{
  colorlinks = true,
  allcolors = inlinkcolor,
  urlcolor = exlinkcolor,
}
\fi

\newenvironment{@abssec}[1]{
        \vspace{.05in}\parindent .0in
        {\upshape\bfseries #1. }\ignorespaces
    }
    {\par\vspace{.1in}}
\renewenvironment{abstract}{\begin{@abssec}{\abstractname}}{\end{@abssec}}
\newenvironment{keywords}{\begin{@abssec}{Keywords}}{\end{@abssec}}

\usepackage{fancyhdr}

\author{
  {\normalsize Qinmeng Zou}\thanks{CentraleSup\'elec, Universit\'e Paris-Saclay, 3 rue Joliot Curie, 91190 Gif-sur-Yvette, France
    (\email{zouqinmeng@gmail.com}, \email{frederic.magoules@hotmail.com}).}
  \and
  {\normalsize Fr\'ed\'eric Magoul\`es\footnotemark[1]}
}
\title{Recent Developments in Iterative Methods for Reducing Synchronization}
\date{}

\begin{document}
\maketitle
\thispagestyle{fancy}

\begin{abstract}
On modern parallel architectures, the cost of synchronization among processors can often dominate the cost of floating-point computation.
Several modifications of the existing methods have been proposed in order to keep the communication cost as low as possible.
This paper aims at providing a brief overview of recent advances in parallel iterative methods for solving large-scale problems.
We refer the reader to the related references for more details on the derivation, implementation, performance, and analysis of these techniques.
\end{abstract}


\section{Introduction}

The performance of iterative methods depends on the amount of arithmetic operations and the amount of data movements.
The latter depends further on the latency cost and the bandwidth cost.
Early researches mainly focused on the arithmetic operations in both the sequential and parallel cases.
On modern computer architectures, latency cost is much more significant than bandwidth cost, and bandwidth cost is much more significant than computation cost.
The gaps are expected to increase in the future.

Parallelization of iterative methods for solving large-scale problems is constrained by synchronization which leads to processor idle time.
These methods consist of the following three basic operations: sparse matrix-vector multiplication (SpMV), dot products, and AXPY ($\alpha$ times $x$ plus $y$) operations
\[
y\leftarrow\alpha x+y.
\]
AXPY requires only local operations and thus does not affect parallel efficiency.
SpMV often requires communication among neighbors, which depends on the distribution of nonzero values.
Dot products require global synchronization before and after a computation.
The bottleneck that comes from these operations can be partially overcome by using recent techniques.
Roughly speaking, the communication in dot products can be reduced by simultaneously constructing multiple direction vectors or being overlapped with other operations; if there is no dot product operation, then the synchronization in SpMV can be eliminated by only using existing data for the next computation instead of waiting for a complete data transmission.

We show in this paper how synchronization points could be reduced and highlight some recent advances on promising techniques.
Section~\ref{sec:2} presents the $s$-step iterative methods.
Section~\ref{sec:3} presents the pipelined Krylov subspace methods.
Section~\ref{sec:4} summarizes recent developments in asynchronous iterations.
Section~\ref{sec:5} provides a brief overview of other popular methods.
Finally, we draw a conclusion in Section~\ref{sec:6}.

\section{s-Step Iterative Methods}\label{sec:2}

An early paper describing the idea of $s$-step iterations can be traced to~1950, as quoted in~\cite{Forsythe1968}, when Birman~\cite{Birman1950} presented an $s$-gradient method in a Russian paper.
A later paper~\cite{Chronopoulos1989} was published on the method that aims at reducing the number of synchronization operations for the conjugate gradient (CG) method~\cite{Hestenes1952}, often called Chronopoulos-Gear CG.
The thesis written by~Hoemmen~\cite{Hoemmen2010} gives an excellent historical perspective on this topic.
We refer the reader to~\cite{Hoemmen2010} and the references therein for the developments before~2010.

The key feature in $s$-step iterative methods is to perform~$\mathcal{O}(s)$ computation steps of the classical algorithms for each communication step, thus allowing to reduce the number of synchronization points by a factor of~$s$.
Krylov subspace methods (KSMs) are often the methods of choice for solving eigenvalue and linear system problems (see,~e.g,~\cite{Golub2013}), which can be viewed as projection techniques.
Given two subspace~$\mathcal{K}_m$ and~$\mathcal{L}_m$ where
\[
\mathcal{K}_m(A,v)=\text{span}\left\{v,\,Av,\,\dots,\,A^{m-1}v\right\},
\]
KSMs search solution vectors in~$\mathcal{K}_m$ such that residuals are orthogonal to~$\mathcal{L}_m$.
The latent bottleneck of successive matrix-vector multiplications in KSMs can be relieved by the ``matrix powers kernel'' as described in~\cite{Demmel2008}.
The global synchronization of dot products in Lanczos-based methods can be reduced by using a Gram matrix~\cite{Carson2013}.
In addition, the GMRES method~\cite{Saad1986} was improved in~\cite{Hoemmen2010} by using the tall and skinny QR factorization~\cite{Demmel2012}.

Hoemmen~\cite{Hoemmen2010} discussed the $s$-step GMRES (see also~\cite{Mohiyuddin2009}) and $s$-step CG.
Carson et al.~\cite{Carson2013} discussed the $s$-step BICG~\cite{Fletcher1976} and $s$-step BICGSTAB~\cite{vanderVorst1992}.
They also addressed the stability issues relating to the basis construction.
The term ``$s$-step'' can be often replaced by ``communication-avoiding (CA)''.
Although the latter is commonly used in the literature, as mentioned in~\cite{Cools2018}, it is slightly dubious since the communication cost is only partially reduced rather than avoided.
Ballard et al.~\cite{Ballard2014} summarized theoretical bounds on communication for techniques used in the $s$-step algorithms.

In finite precision cases, $s$-step formulation with monomial basis can lead to stability issues, which have been discussed in many references~\cite{Joubert1992,Bai1994,Hoemmen2010,Carson2013}.
The maximum attainable accuracy of $s$-step KSMs and the residual replacement strategy were discussed in~\cite{Carson2014}.

More recently, a new variant proposed by Imberti and Erhel~\cite{Imberti2017} is to use an increasing sequence of block sizes in $s$-step GMRES instead of a fixed size.
On the other hand, block coordinate descent (BCD) methods have been successfully used in machine learning.
Devarakonda et al.~\cite{Devarakonda2019} extended the $s$-step methods to the primal and dual BCD methods for solving regularized least-square problems.
The new methods are called CA-BCD and CA-BDCD, respectively, that can, like other $s$-step iterative methods, reduce the latency cost by a factor of~$s$ but increase computation and bandwidth costs.

\section{Pipelined Krylov Subspace Methods}\label{sec:3}

The pipelined iterative methods aim at overlapping expensive communication phases with computations.
Some approaches for this purpose applied to CG and GMRES appeared in the mid-1990s, see~\cite{Demmel1993,deSturler1995}.
Ghysels et al.~\cite{Ghysels2013,Ghysels2014} introduced the modern pipelined Krylov subspace methods.
It is interesting to note that the term ``pipelined'' has often been replaced by ``communication-hiding'' in their camp, just like the alternatives mentioned in the preceding section.

Ghysels et al.~\cite{Ghysels2013} proposed the pipelined GMRES method.
Ghysels and Vanroose~\cite{Ghysels2014} proposed the pipelined CG method.
Their work has promoted other promising ideas.
For example, Sanan et al.~\cite{Sanan2016} discussed some pipelined variants of flexible Krylov subspace methods.
Cools and Vanroose~\cite{Cools2017} presented a general framework for pipelined methods, from which the pipelined BICGSTAB method was successfully derived.

In finite precision arithmetic, Cools et al.~\cite{Cools2018} discussed the the effect of local rounding error propagation on the maximal attainable accuracy of the pipelined CG method and compared it with the classical CG and Chronopoulos-Gear CG~\cite{Chronopoulos1989}.
In a later paper, Cools~\cite{Cools2019} gave a similar discussion for the pipelined BICGSTAB method.
Carson et al.~\cite{Carson2018b} discussed the stability issues for synchronization-reducing algorithms and presented a methodology for the theoretical analysis of some CG variants.

\section{Asynchronous Iterations}\label{sec:4}

Asynchronous iterations were introduced by Chazan and Miranker~\cite{Chazan1969} in~1969, originally called chaotic iterations.
The modern mathematical expression was formally introduced in~\cite{Baudet1978}, which can be summarized as follows:
\[
x_i^{(n+1)} =
\begin{cases}
f_i\left(x_1^{(\tau_{i,1,n})},\,\dots,\,x_p^{(\tau_{i,p,n})}\right), & i\in P_n, \\
x_i^{(n)}, & i\notin P_n,
\end{cases}
\]
where $x_i^{(n)}$ denotes the $i$th element of the solution vector, $\tau_{i,j,n}$ denotes the iteration number with retards for each element $j$ in each processor $i$, which is smaller than~$n$, and $P_n\subset\{1,\,\dots,\,p\}$ is a subset of processors.
In addition, some multi-stage models and a number of convergence results have been proposed during the past century, we refer the reader to~\cite{Frommer2000,Bahi2007} for more details.
The key feature in asynchronous iterations is to eliminate waiting times in communication at the expense of more iterations.

The main focus of recent research is asynchronous domain decomposition methods.
Chau et al.~\cite{Chau2014} used the asynchronous Schwarz method for the solution of obstacle problems.
Magoul\`es et al.~\cite{Magoules2017} investigated the asynchronous optimized Schwarz method and provided some convergence results.
Magoul\`es and Venet~\cite{Magoules2018c} gave a discussion about the asynchronous substructuring methods.
More recently, Yamazaki et al.~\cite{Yamazaki2019} gave some numerical experiments for the optimized Schwarz method.
Theoretical analysis of this approach for various types of subdomains were considered in~\cite{Garay2018,Garay2018b}.
On the other hand, asynchronous multisplitting methods were studied in~\cite{Bahi1997}, which have been applied to the fluid-structure interaction problem in a recent paper~\cite{Partimbene2019}.

For the time domain decomposition methods, the derivation of asynchronous waveform relaxation can be found in~\cite{Frommer2000}.
Magoul\`es et al.~\cite{Magoules2018,Magoules2018d} proposed the asynchronous variant of Parareal algorithm (see also~\cite{Zou2017,Zou2017b}).
Another asynchronous time-parallel method based on Laplace transform can be found in~\cite{MagoulesX1}.

From a computational point of view, the implementation of asynchronous iterative methods requires more than a straightforward update of synchronous versions.
Magoul\`es and Gbikpi-Benissan~\cite{Magoules2017b,Magoules2018b} developed an MPI-based communication library for both synchronous and asynchronous iterative computing.
However, the issue of asynchronous convergence detection must be tackled for such kind of libraries.
Early work can be found in~\cite{Savari1996} based on the snapshot algorithm.
Magoul\`es and Gbikpi-Benissan~\cite{Magoules2018e} continued this work and proposed several promising variants.
Bahi et al.~\cite{Bahi2005} introduced another approach in which the detection process is superimposed onto the asynchronous iterations.
Numerical experiments of asynchronous iterative methods for GPU were conducted and described in~\cite{Anzt2013,Chow2015}.

\section{Other Popular Methods}\label{sec:5}

It is clear that other techniques exist for reducing synchronization in parallel architectures, which have not been reviewed in the previous sections.
For example, cyclic gradient iterative methods (see,~e.g.,~\cite{ZouX1}) are intrinsically adapted for parallel computing, which can reduce both computation and communication costs.
Some techniques (see,~e.g.~\cite{Gu2007}) called improved Krylov methods revolve around the overlap of communication and computation.
McInnes et al.~\cite{McInnes2014} proposed hierarchical and nested Krylov subspace methods.
Grigori et al.~\cite{Grigori2016} proposed enlarged Krylov subspace methods which can be viewed as a special case of augmented Krylov subspace methods (see,~e.g,~\cite{Simoncini2007}).

\section{Conclusion}\label{sec:6}

There is much still to understand about synchronization-reducing methods.
For example, development of efficient preconditioners for parallel algorithms is still an open question~\cite{Carson2015b}.
Theoretical analysis of basis in $s$-step algorithms requires more work~\cite{Hoemmen2010}.
The loss of orthogonality in finite precision is also an issue to be tackled~\cite{Carson2018b}.
For this reason, we hope that continued contributions could be made on this rapidly growing field in the future.

\section*{Acknowledgment}

This work was funded by the project ADOM (M\'ethodes de d\'ecomposition de domaine asynchrones) of the French National Research Agency (ANR).

\bibliography{ref}
\bibliographystyle{abbrv}

\end{document}